\documentclass[twocolumn,preprintnumbers,amsmath,amssymb,pra]{revtex4}

\usepackage[dvips]{graphicx}
\usepackage{dcolumn}
\usepackage[normalem]{ulem}
\usepackage{subfigure}
\usepackage{amsmath}
\usepackage{amssymb}
\usepackage{upgreek}
\usepackage{psfrag}
\usepackage{color}

\newcommand{\bm}[1]{\boldsymbol{\mathbf{#1}}}
\newcommand{\ud}{\mathrm{d}}
\newcommand{\e}{\bm{E}}

\newcommand{\tens}[1]{\bm{#1}}
\newcommand{\re}{\operatorname{Re}}
\newcommand{\im}{\operatorname{Im}}
\newcommand{\tr}{\operatorname{Tr}}

\begin{document}

\preprint{}

\title{Absorption by an optical dipole antenna in a structured environment}

\author{E. Castani\'e}
\author{R. Vincent}
\author{R. Pierrat}
\email{romain.pierrat@espci.fr}
\author{R. Carminati}
\email{remi.carminati@espci.fr}
\affiliation{Institut Langevin, ESPCI ParisTech, CNRS, 10 rue Vauquelin, 75231 Paris Cedex 05, France}

\date{\today}

\begin{abstract}
   We compute generalized absorption and extinction cross-sections of an optical dipole nanoantenna in a
   structured environment. The expressions explicitly show the influence of radiation reaction and the local density of states
   on the intrinsic absorption properties of the antenna. Engineering the environment could allow to modify
   the overall absorption as well as the frequency and the linewidth of a resonant antenna. 
   Conversely, a dipole antenna can be used to probe the photonic environment, in a similar way as
   a quantum emitter.
\end{abstract}

\maketitle

\section{Introduction}

It is well-known that the emission frequency and linewidth of a dipole quantum emitter is modified by its
local environment~\cite{PURCELL-1946,CHANCE-1978,BARNES-1998,NOVOTNY-2006}. The linewidth directly depends on the
photonic Local Density Of States (LDOS) which accounts for the number of radiative and non-radiative channels available for the emitter
to relax in the ground state. The change in the emission frequency and linewidth induce by the environment can be described
by considering the transition dipole as a classical dipole oscillator~\cite{CHANCE-1978,NOVOTNY-2006}.
Therefore similar behaviors are expected for a dipole antenna (or a nanoparticle) interacting with its environment, 
the involved dipole being in this case the {\it induced} dipole that is responsible for scattering and absorption.
Indeed changes in the induced dipole dynamics in optical antennas have already been 
observed~{\cite{HALL-1984,SANDOGHDAR-2005,SANDOGHDAR-2005-1}, and the parallel with spontaneous emission dynamics
has been mentioned on a qualitative phenomenological ground. Energy shifts and linewidths
of plasmonic nanoparticles have also been discussed recently, based on general properties of damped harmonic
oscillators~\cite{NORDLANDER-2011}. A connection between LDOS maps and
the scattering pattern of plasmonic structures has been established in a specific imaging
configuration~\cite{HUANG-2008-1}. In this context, it seems that a general discussion of
the influence of the environment on the absorption of an optical dipole antenna or nanoparticle would be useful. 
The purpose of this paper is to address this question, using a rigorous framework based on scattering theory, and
to illustrate the conclusions on a simple example.

In this work, we investigate the role of the environment on the absorption cross-section of an optical dipole antenna
using a rigorous theoretical framework. We show that it is possible to modify the overall absorption and its
spectral properties by engineering the environment, and in particular the photonic LDOS. 
Conversely, it is possible to probe the environment using a resonant nanoantenna, specific measurements being able
to produce LDOS maps.
In section II and III, we derive the exact expressions of the dressed electric polarizability of a dipole antenna
in an arbitrary environment, and deduce the expression of the generalized absorption and extinction cross-sections. 
In section IV, we discuss qualitatively the physical mechanisms affecting both the resonance frequency and the linewidth
of a resonant antenna, using a simplified model. In section V we study numerically a simple but realistic example,
based on the rigorous expression derived in section II. This allows us to illustrate the general trends and to give orders of magnitude.
In section VI we summarize the main conclusions.

\section{Dressed polarizability}

\begin{figure}[!htbf]
   \centering
   \psfrag{r}[c]{$\bm{r}_0$}
   \psfrag{a}[c]{$\tens{\alpha}(\omega)$}
   \psfrag{E}[c]{$\e_{\textrm{inc}}$}
   \includegraphics[width=0.8\linewidth]{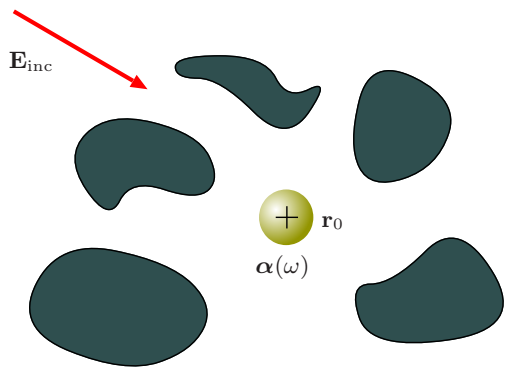}
   \caption{Optical dipole antenna of arbitrary shape in a structured environment. 
   A point $\bm{r}_0$ inside the antenna is used to define its position.}.
   \label{system}
\end{figure}

This section is devoted to the computation of the dressed polarizability of an optical dipole antenna or nanoparticle (i.e. the polarizability that accounts
for the interaction with the environment). To proceed, we follow the same procedure that has been used previsouly to compute the
polarizability in vacuum~\cite{LAKHTAKIA-1992,CARMINATI-2006,CARMINATI-2010-2}.
We consider an electrically small particle (the generic term particle will
be used to denote either a subwavelength optical antenna or a nanoparticle) of volume $V$ and of permittivity $\epsilon(\omega)$
embedded in an arbitrary environment, the particle lying at position $\bm{r}_0$ (see figure~\ref{system}).
We assume that position $\bm{r}_0$ lies in vacuum (the local refractive index at $\bm{r}_0$ is assumed to be unity).
To describe light
propagation  in  the environment, we use the electric dyadic Green function $\tens{G}$ which connects the electric field at position $\bm{r}$
to an electric dipole source at point $\bm{r}'$ through the relation $\e(\bm{r})=\mu_0\omega^2\tens{G}(\bm{r},\bm{r}')\bm{p}(\bm{r})$. We denote
by $\e_{\textrm{ext}}$ the field in the environment in the absence of the particle (exciting field). The total
electric field $\e$ at point $\bm{r}$ and at frequency $\omega$ reads
\begin{align}\nonumber
   \e(\bm{r},\omega)= \e_{\textrm{ext}}(\bm{r},\omega)+k_0^2\int_V & \tens{G}(\bm{r},\bm{r}',\omega)[\epsilon(\omega)-1]
\\\label{field} & \times
   \e(\bm{r}',\omega)\ud^3\bm{r}'
\end{align}
where $k_0=\omega/c$, $c$ being the speed of light in vacuum.
The approximation of electrically small particle amounts to considering that the electric field is uniform inside the particle. 
Under this condition, the expression of the total electric field {\it inside} the particle (at point $\bm{r}_0$) becomes
\begin{align}\nonumber
   \e(\bm{r}_0,\omega)= \e_{\textrm{ext}}(\bm{r}_0,\omega)+k_0^2[\epsilon(\omega)-1] \e(\bm{r}_0,\omega)
\\\label{field2}  \times
   \int_V \tens{G}(\bm{r}_0,\bm{r}',\omega)\ud^3\bm{r}' .
\end{align}
We now split the integral of the Green dyadic into its singular part $-\tens{L}/k_0^2$ and its non-singular part
$V\tens{G}_{\textrm{reg}}(\bm{r}_0,\bm{r}_0,\omega)$ where we have assumed that $\tens{G}_{\textrm{reg}}(\bm{r}_0,\bm{r}',\omega)$ is
constant over the volume of the nanoparticle. Note that $\tens{L}$ is real since it corresponds to the singularity of the Green tensor
at a point that lies in vacuum (the Green tensor is computed in the absence of the particle)~\cite{BLADEL-1991,YAGHJIAN-1980,GUERIN-2007-1}. 
Equation~(\ref{field2}) becomes
\begin{align}\nonumber
   \e(\bm{r}_0,\omega)= & \left\{
      \tens{I}+[\epsilon(\omega)-1]\tens{L}
   \right.
\\\label{field3} &
   \left.
      -k_0^2V[\epsilon(\omega)-1]\tens{G}_{\textrm{reg}}(\bm{r}_0,\bm{r}_0,\omega)
   \right\}^{-1}
   \e_{\textrm{ext}}(\bm{r}_0,\omega).
\end{align}
The expression of the polarizability follows be writing the induced electric dipole moment of the particle in the form
\begin{align}\nonumber
   \bm{p}(\bm{r}_0,\omega) & =\int_V \bm{P}(\bm{r}',\omega)\ud^3\bm{r}'=V\epsilon_0[\epsilon(\omega)-1]\e(\bm{r}_0,\omega)
\\\label{dipole} & 
   =\epsilon_0\tens{\alpha}(\omega)\e_{\textrm{ext}}(\bm{r}_0,\omega).
\end{align}
The last line defines the dressed polarizability $\tens{\alpha}(\omega)$, that in the most general situation is a tensor.
Inserting Eq.~(\ref{field3}) into Eq.~(\ref{dipole}), one obtains
\begin{align}\nonumber
   \tens{\alpha}= & V[\epsilon(\omega)-1]
\\\label{pola} & \times
   \left\{
      \tens{I}+[\epsilon(\omega)-1]\tens{L}-k_0^2V[\epsilon(\omega)-1]\tens{G}_{\textrm{reg}}(\bm{r}_0,\bm{r}_0,\omega)
   \right\}^{-1}.
\end{align}
A more useful expression of the dressed polarizability is obtained by defining a reference polarizablity $\tens{\alpha}_0(\omega)$.
A usual choice for this reference is the quasi-static polarizability of the particle in vacuum~\cite{CARMINATI-2010-2}, that reads
\begin{align}\label{pola0}
   \tens{\alpha}_0(\omega)=V[\epsilon(\omega)-1]\left\{\tens{I}+[\epsilon(\omega)-1]\tens{L}\right\}^{-1}.
\end{align}
Note that in the case of a spherical particle, the singularity (or depolarization) dyadic is $\tens{L}=\tens{I}/3$
so that  $\tens{\alpha}_0(\omega)$ would simplify into the well-known (scalar) quasi-static expression $\alpha_0=3V(\epsilon-1)/(\epsilon+2)$. 
Using Eqs.~(\ref{pola})
and~(\ref{pola0}), the dressed polarizability has the final form
\begin{align}\label{pola_final}
   \tens{\alpha}(\omega)=\tens{\alpha}_0(\omega)
      \left\{\tens{I}-k_0^2\tens{G}_{\textrm{reg}}(\bm{r}_0,\bm{r}_0,\omega)\tens{\alpha}_0(\omega)\right\}^{-1}.
\end{align}
This expression is the main result of this section. It shows that the dressed polarizability of a particle depends on the
environment, the influence of the environment being fully described by the non-singular part of the dyadic Green function
$V\tens{G}_{\textrm{reg}}(\bm{r}_0,\bm{r}_0,\omega)$. Note that Eq.~(\ref{pola_final}) is the explicit expression
of the effective polarizability discussed in Ref.~\cite{NOVOTNY-2006}.

For a resonant antenna or nanoparticle (e.g., supporting a plasmon resonance), it is instructive to rewrite Eq.~\ref{pola_final} in the form
\begin{align}\label{pola_final2}
   \tens{\alpha}(\omega)^{-1}=\tens{\alpha}_0(\omega)^{-1}
     -k_0^2\tens{G}_{\textrm{reg}}(\bm{r}_0,\bm{r}_0,\omega).
\end{align}
The resonance frequency of the dressed polarizability is solution of the equation
$\re[\tens{\alpha}_0^{-1}(\omega)-k_0^2 \tens{G}_{\textrm{reg}}(\bm{r}_0,\bm{r}_0,\omega)]=0$, while the linewidth is given by
$\im[\tens{\alpha}_0^{-1}(\omega)-k_0^2\tens{G}_{\textrm{reg}}(\bm{r}_0,\bm{r}_0,\omega)]$. The influence of the 
environment on the resonance lineshape is made explicit by this simple analysis.

\section{Generalized absorption and extinction cross-sections}

The expression of the dressed polarizability is the starting point to compute generalized absorption and
extinction cross-sections in an arbitrary environment. To carry out this derivation, we start with the expression of the
time-averaged power absorbed inside the particle, given by
\begin{align}
   P_a=\frac{1}{2}\int_V \re[\bm{j}(\bm{r}',\omega)\cdot\e^*(\bm{r}',\omega)]\ud^3\bm{r}'
\end{align}
where $\bm{j}(\bm{r}',\omega)=-i\omega\epsilon_0[\epsilon(\omega)-1]\e(\bm{r}',\omega)$
is the current density induced in the particle.
 As the electric field is assumed
to be uniform inside the particle, the absorbed power becomes
\begin{align}
   P_a=\frac{\omega V\epsilon_0 \im\epsilon(\omega)}{2}|\e(\bm{r}_0,\omega)|^2.
\end{align}
Using Eq.~(\ref{dipole}), it can be rewritten as
\begin{align}
   P_a=\frac{\omega\epsilon_0 \im\epsilon(\omega)}{2V|\epsilon(\omega)-1|^2}|\tens{\alpha}(\omega)\e_{\textrm{ext}}(\bm{r}_0,\omega)|^2.
\end{align}
To define a generalized absorption cross-section, we have to introduce an incident local energy  flux $\phi_\textrm{ext}$. Since the exciting
field at the position of the particle is $\e_{\textrm{ext}}(\bm{r}_0,\omega)$, we define the incident local energy  flux using the expression
for a plane wave $\phi_{\textrm{ext}}=|\e_{\textrm{ext}}(\bm{r}_0,\omega)|^2/(2\mu_0c)$. This definition is arbitrary, but has
the advantage to coincide with the standard one when the particle lies in a homogeneous medium. The generalized absorption
cross-section is then given by the ratio $P_a/\phi_{\textrm{ext}}$. Using the relation
$\im[\tens{\alpha}_0][\tens{\alpha}_0\tens{\alpha}_0^*]^{-1}=\im\epsilon(\omega)/\left[V|\epsilon(\omega)-1|^2\right]\tens{I}$
(see appendix~\ref{proof} for a proof), we obtain the final expression 
of the generalized absorption cross-section 
\begin{align}\label{absorption}
   \sigma_a(\omega)\tens{I}=k_0\im[\tens{\alpha}_0(\omega)][\tens{\alpha}_0(\omega)\tens{\alpha}_0^*(\omega)]^{-1}
      \frac{|\tens{\alpha}(\omega)\e_{\textrm{ext}}(\bm{r}_0,\omega)|^2}{|\e_{\textrm{ext}}(\bm{r}_0,\omega)|^2}.
\end{align}
Equation~(\ref{absorption}) is a central result of this paper. It deserves some remarks before we analyze its consequences. 
Although Eq.~(\ref{absorption}) involves tensor notations, the absorption cross-section $\sigma_a(\omega)$ is a scalar quantity.
Expression (\ref{absorption}) is exact and has been obtained
under the only assumption that the electric field is uniform inside the particle (approximation of electrically small particle). It can be used, 
together with Eq.~(\ref{pola_final}), to discuss the influence of the environment on the optical properties of any particle (or antenna) 
satisfying this condition. For a non-absorbing material, the imaginary part
of the permittivity $\epsilon(\omega)$ vanishes, and the quasi-static polarizability $\tens{\alpha}_0$ is real. 
The term $\im[\tens{\alpha}_0(\omega)]$ in Eq.~(\ref{absorption}) implies $ \sigma_a(\omega)=0$, as it should be.
Finally let us emphasize that the generalized absorption cross-section that we have defined
really describes the change of the intrinsic absorption of the particle induced by the environment (or in other word
of the absorption probability given a local incident power). Its proper normalization
by the local incident energy flux clearly distinguishes this effect on the absorbed power from that due to a mere change of the
local incident power.
It is also important to stress that although the generalized absorption cross-section that we have defined
is a scalar, it depends on the orientation of the local exciting electric field, that encodes the anisotropy of the environment.

Following the same procedure, it is also possible to compute the extinction cross-section, 
starting from the expression of the power extracted from the external field by
the nanoparticle. The latter can be written in the form~\cite{CARMINATI-2010-2}:
\begin{align}
   P_e=\frac{1}{2}\int_V \re[\bm{j}(\bm{r}',\omega)\cdot\e_{\textrm{ext}}^*(\bm{r}',\omega)]\ud^3\bm{r}'.
\end{align}
One obtains
\begin{align}\label{extinction}
   \sigma_e=k_0\frac{\im[\tens{\alpha}(\omega)\e_{\textrm{ext}}(\bm{r}_0,\omega)\cdot\e_{\textrm{ext}}^*(\bm{r}_0,\omega)]}
      {|\e_{\textrm{ext}}(\bm{r}_0,\omega)|^2}.
\end{align}
As for the generalized absorption cross-section, this expression is exact under the assumption of an electrically
small particle. In the following, we will focus our attention on the absorption cross-section, but the analyses and
the general trends can be translated to the extinction situation, that can be relevant to specific experimental configurations
and type of measurements.
 
\section{Qualitative discussion}

Equation~(\ref{absorption}) can be used to compute the absorption cross-sections in a given environment. For realistic geometries,
the computation can only be performed numerically (we will study a simple example in section V).
For example, it is possible to use an iteration scheme to solve numerically the Dyson equation which is the closed form
of the equation governing the Green function similar to Eq.~(\ref{field})~\cite{DEREUX-1995}.
Nevertheless, in order to get some insight based on simple analytical formulas, we will study an oversimplified
situation. First, we consider a spherical nanoparticle of volume $V$, small enough to be consistent with the electric dipole approximation that
we use throughout this work. For such a shape, the singular part of the Green tensor is simply given by
 $\tens{L}=\tens{I}/3$~\cite{BLADEL-1991,YAGHJIAN-1980}. Second, we assume that
the nanosphere is embedded in an environment that preserves for $\tens{G}_{\textrm{reg}}(\bm{r}_0,\bm{r}_0,\omega)$ the same symmetry
as that of free-space [i.e. $\tens{G}_{\textrm{reg}}(\bm{r}_0,\bm{r}_0,\omega)=G_{\textrm{reg}}(\bm{r}_0,\bm{r}_0,\omega)\tens{I}$].
This is an unrealistic hypothesis (it would be strictly valid only for a homogeneous medium or a medium with cubic symmetry) 
but we shall use it only to discuss qualitatively general trends. Under these hypotheses, the quasi-static polarizability reduces to
$\alpha_0(\omega)=3V[\epsilon(\omega)-1]/[\epsilon(\omega)+2]$, and the dressed polarizability
takes the form
\begin{align}
   \alpha(\omega)=\frac{\alpha_0(\omega)}{1-k_0^2G_{\textrm{reg}}(\bm{r}_0,\bm{r}_0,\omega)\alpha_0(\omega)}.
\end{align}
The absorption cross-section becomes
\begin{align}
   \sigma_a(\omega)=k_0\im[\alpha_0(\omega)]\frac{|\alpha(\omega)|^2}{|\alpha_0(\omega)|^2}.
\end{align}
It is clear in this expression that in absence of polarization anisotropy induced by the environment, the absorption cross-section does not depend on the
exciting field [this is not the case in Eq.~(\ref{absorption})]. 

In order to get a simple model of a resonant optical dipole antenna, we consider a metallic nanoparticle described by a Drude permittivity
$\epsilon(\omega)=1-\omega_p^2/(\omega^2+i\omega\gamma)$ where $\omega_p$ is the plasma frequency and $\gamma$ the intrinsic 
collision rate that describes absorption losses in the bulk material. 
The real and imaginary parts of the Green tensor, that both influence the absorption cross-section, have a well-defined meaning.
In order to makes this more explicit, we introduce the photonic LDOS $\rho(\bm{r}_0,\omega)$, connected to the imaginary part of the Green tensor by
\begin{align}
   \rho(\bm{r}_0,\omega)=\frac{6\omega}{\pi c^2}\im[G_{\textrm{reg}}(\bm{r}_0,\bm{r}_0,\omega)].
\end{align}
We also introduce $\phi(\bm{r}_0,\omega)$ that describes the influence of the real part of the non-singular Green tensor
in a similar way:
\begin{align}
   \phi(\bm{r}_0,\omega)=\frac{6\omega}{\pi c^2}\re[G_{\textrm{reg}}(\bm{r}_0,\bm{r}_0,\omega)].
\end{align}
Using these definitions, the dressed polarizability of the metallic nanoparticle reads
\begin{align}\nonumber
   \alpha(\omega)=3V\omega_0^2 &
   \left[\omega_0^2\left\{1-\frac{\pi}{2}V\omega\phi(\bm{r}_0,\omega)\right\}\right.
\\\label{pola_final_scalaire} & 
   \left.-\omega^2-i\omega\left\{\gamma+\frac{\pi}{2} V\omega_0^2\rho(\bm{r}_0,\omega)\right\}\right]^{-1}
\end{align}
where $\omega_0=\omega_p/\sqrt{3}$ is the plasmon resonance frequency of the bare nanoparticle in the quasi-static limit. 
This expression naturally leads to the introduction of an effective frequency
$\Omega_{\textrm{eff}}^2(\omega)=\omega_0^2\left\{1-\pi V\omega\phi(\bm{r}_0,\omega)/2\right\}$
such that the resonant frequency of the particle in the environment is solution of the equation
$\Omega_{\textrm{eff}}^2(\omega)-\omega^2=0$. Similarly, an effective linewidth
$\gamma_{\textrm{eff}}(\omega)=\gamma+\pi V\omega_0^2\rho(\bm{r}_0,\omega)/2$ can be introduced. 

From Eqs.~(\ref{absorption}) and~(\ref{pola_final_scalaire}), we can obtain the expression of the absorption cross-section 
in the simplified scalar model:
\begin{align}\label{absorption_scalaire}
   \sigma_a(\omega)=\frac{3V\omega_0^2}{c}\frac{\omega^2\gamma}
      {\left[\Omega_{\textrm{eff}}^2(\omega)-\omega^2\right]^2+\omega^2\gamma_{\textrm{eff}}^2(\omega)}.
\end{align}
Equation~(\ref{absorption_scalaire}), together with the expressions of $\Omega_{\textrm{eff}}(\omega)$ and 
$\gamma_{\textrm{eff}}(\omega)$, shows that the real part of the Green function contributes to a change
of the resonance frequency, while the imaginary part (the LDOS) changes the linewidth. This is the same
behavior as that known for a dipole emitter (either quantum or classical), although in the present situation
we deal with the dipole induced inside the particle by the external field.  
It is interesting to note that the effective linewidth $\gamma_{\textrm{eff}}(\omega)$ can only be
larger than the intrinsic linewidth $\gamma$ because of its dependance on the LDOS which is a positive quantity.

The resonant behavior of $\sigma_a(\omega)$ and the influence of the LDOS [through $\gamma_{\textrm{eff}}(\omega)$] 
deserve to be analyzed more precisely.
Due to the frequency dependence of both $\Omega_{\textrm{eff}}(\omega)$ and $\gamma_{\textrm{eff}}(\omega)$,
the resonance lineshape is in general not a Lorentzian profile. Moreover, $\Omega_{\textrm{eff}}(\omega)$ and 
$\gamma_{\textrm{eff}}(\omega)$ are not independent, since the real and imaginary parts
of the Green function are connected by Kramers-Kronig relations.
It is nevertheless possible to derive the expression of the generalized absorption cross-section at resonance.
The resonance frequency $\omega_a$ satisfies $\ud\sigma_a(\omega_a)/\ud\omega=0$.
As described in appendix~\ref{resonant},
using this implicit equation,
it is possible to express $\sigma_a(\omega_a)$ in the form
\begin{align}\nonumber
   \sigma_a & (\omega_a) = \frac{3\gamma V\omega_0^2}{c} \left[\gamma_{\textrm{eff}}^2(\omega_a)\left\{
      1
   \vphantom{\frac{\omega_a^4}{\omega_a^4}}
   \right.\right.
\\\label{absorption_resonant} &
   \left.\left.
      +\frac{\omega_a^4}{[\Omega_{\textrm{eff}}^2(\omega_a)-2\omega_a\Omega_{\textrm{eff}}'(\omega_a)\Omega_{\textrm{eff}}(\omega_a)
      +\omega_a^2]^2}\gamma_{\textrm{eff}}'^2(\omega_a)
   \right\}\right]^{-1}.
\end{align}
where the superscript $\prime$ denotes a first-order derivative.
This expression shows that both the LDOS and its first-order derivative influence the amplitude of the generalized
absorption cross-section, through $\gamma_{\textrm{eff}}^2(\omega_a)$ and $\gamma_{\textrm{eff}}'^2(\omega_a)$,
respectively. An increase of both quantities tends to decrease the absorption cross-section. In the particular
case of an environment for which the spectral dependence of the LDOS can be neglected [$\gamma_{\textrm{eff}}'^2(\omega) = 0$],
we end up with $\sigma_a(\omega_a) \propto 1/\gamma_{\textrm{eff}}^2(\omega_a)$. This result can be
qualitatively explained in simple terms.
The first step in the absorption process by a metallic nanoparticle is the excitation of the conduction electron gas.
Then relaxation can occur either by radiative (emission of scattered light) or non-radiative channels
(absorption due to electron-phonon collisions). Increasing the photonic LDOS increases the weight of radiative channels and
therefore reduces absorption. The role of the LDOS in this process is essentially the same as that in the spontaneous decay rate of a
quantum emitter by coupling to radiation.

\section{Metallic nanoparticle interacting with a perfect mirror}

\begin{figure}[!htbf]
   \centering
   \psfrag{R}[c]{$-\bm{r}_0$}
   \psfrag{r}[c]{$\bm{r}_0$}
   \psfrag{x}[c]{$x$}
   \psfrag{y}[c]{$y$}
   \psfrag{z}[c]{$z$}
   \psfrag{d}[c]{$d$}
   \psfrag{t}[c]{$\theta$}
   \psfrag{E}[c]{$\e_{\textrm{inc}}$}
   \includegraphics[width=0.6\linewidth]{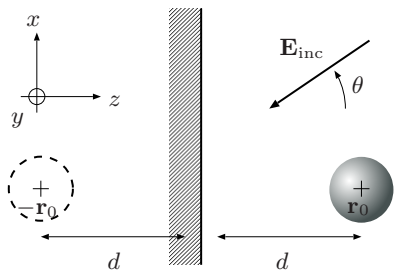}
   \caption{Geometry of the system: A spherical dipole nanoparticle interacts with a perfect mirror. The dashed sphere
   corresponds to the image dipole. }
   \label{mirror}
\end{figure}

In order to illustrate the effects discussed above on a real example, and to get orders of magnitudes
(i.e. to establish the possibility of experiments),
we study quantitaively in this section the generalized absorption cross-section of a silver nanosphere,
with radius $R=15\,\textrm{nm}$, interacting with a flat
perfectly conducting surface (perfect mirror).
The geometry of the system is shown in Fig.~\ref{mirror}. To describe the silver nanoparticle, 
we use the tabulated values of the bulk permittivity $\epsilon(\omega)$ taken from Ref.~\cite{PALIK-1985}.

To compute the generalized absorption cross-section of the nanosphere, we use the exact expression Eq.~(\ref{absorption}).
In order to compute relative changes, we define the normalized cross-section
$\sigma_a^n(\omega)=\sigma_a(\omega)/\sigma_a^{\textrm{vac}}(\omega)$, where $\sigma_a^{\textrm{vac}}(\omega)$
is the absorption cross-section of the bare nanosphere in vacuum:
\begin{align}\label{absorption_vac}
   \sigma_a^{\textrm{vac}}(\omega)=k_0\im[\alpha_0(\omega)]
      \left|\frac{\alpha_{\textrm{vac}}(\omega)}{\alpha_0(\omega)}\right|^2.
\end{align}
The vacuum polarizability $\alpha_{\textrm{vac}}(\omega)$ is given by
\begin{align}\label{pola_final_vac}
   \alpha_{\textrm{vac}}=\alpha_0(\omega)\left\{1-i\frac{k_0^3}{6\pi}\alpha_0(\omega)\right\}^{-1}
\end{align}
with $\alpha_0(\omega)=3V[\epsilon(\omega)-1]/[\epsilon(\omega)+2]$~\cite{CARMINATI-2006,CARMINATI-2010-2}. 
The calculation of $\sigma_a^n(\omega)$ requires the calculation of the exciting field $\e_{\textrm{ext}}$ 
and of the Green function $\tens{G}_{\textrm{reg}}(\bm{r}_0,\bm{r}_0,\omega)$ in the geometry in Fig.~\ref{mirror}. 
This is a straightforward application of the image method, given in appendix~\ref{computation} for completeness. 

\begin{figure*}[!htbf]
   \centering
   \psfrag{d}[c]{$d\,(\mathrm{nm})$}
   \psfrag{l}[c]{$\lambda\,(\mathrm{nm})$}
   \subfigure[]{\label{ldos}\includegraphics[width=0.45\linewidth]{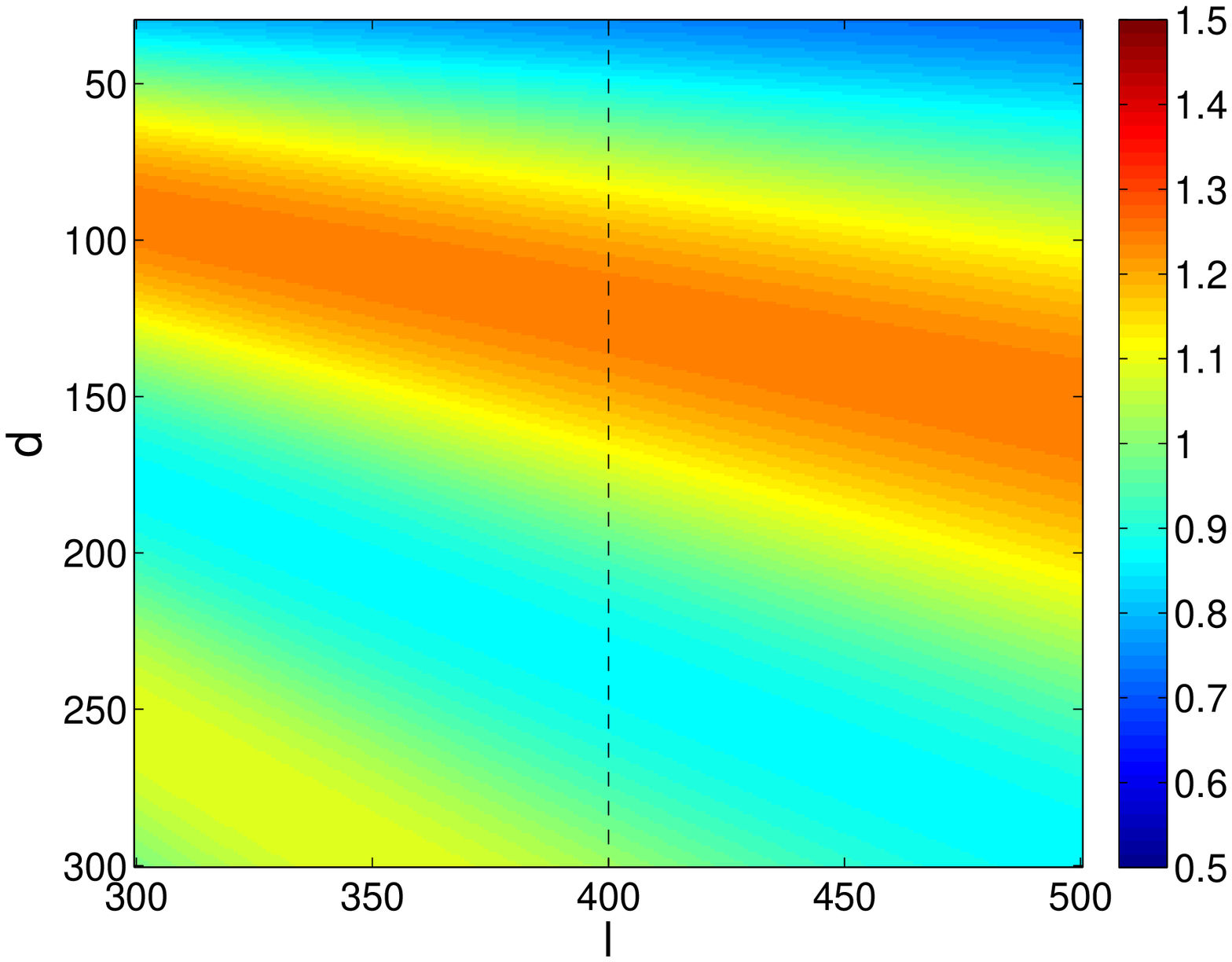}}
   \quad
   \psfrag{d}[c]{$d\,(\mathrm{nm})$}
   \psfrag{Lh}[Bl]{$\rho(\bm{r}_0,\omega)$}
   \psfrag{Lxh}[Bl]{$\rho_{xx,yy}(\bm{r}_0,\omega)$}
   \psfrag{Lzh}[Bl]{$\rho_{zz}(\bm{r}_0,\omega)$}
   \subfigure[]{\label{ldos_partielle}\includegraphics[width=0.425\linewidth]{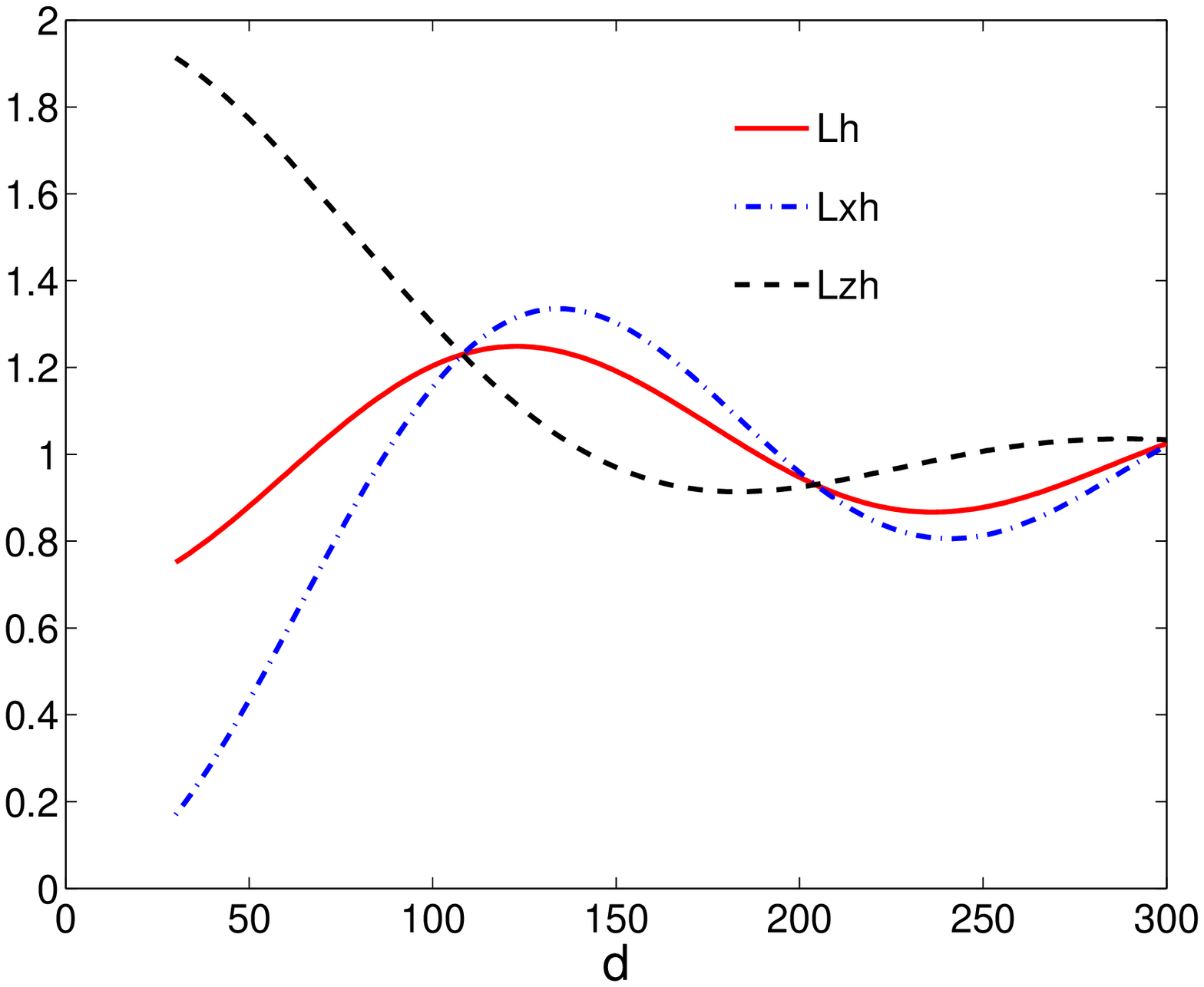}}
   \caption{\subref{ldos} Map of the normalized LDOS $\rho(\bm{r}_0,\omega)/\rho_{\textrm{vac}}(\omega)$,
   with $\rho(\bm{r}_0,\omega)=2\omega/(\pi c^2)\im\left[\tr\tens{G}_{\textrm{reg}}(\bm{r}_0,\bm{r}_0,\omega)\right]$ 
   and $\rho_{\textrm{vac}}(\omega)$ the LDOS in vacuum, versus both the wavelength $\lambda$
   and the distance $d$ between the mirror and the center of the nanoparticle.
   The vertical dashed line is a guide to precise the wavelength at which panel~\subref{ldos_partielle} has been plotted.
   \subref{ldos_partielle} Distance dependence of the LDOS (red solid line) and of the partial LDOS (blue and black dashed lines) defined by
   $\rho_{ii}(\bm{r}_0,\omega)=2\omega/(\pi c^2)\im\left[G_{\textrm{reg},ii}(\bm{r}_0,\bm{r}_0,\omega)\right]$ with $i=x,y,z$
   for $\lambda=400\,\textrm{nm}$. The partial LDOS $\rho_{ii}(\bm{r}_0,\omega)$ drives the dynamics of a dipole oriented
   along direction $i$, while the LDOS sums up the contributions of the three orientations. Due to symmetry in the geometry
   in Fig.~\ref{mirror}, the partial LDOS along $x$ and $y$ are equal.
   In all calculations, the minimum distance $d \simeq 2R$ has been chosen at the limit of validity of the dipole 
   approximation~\cite{CHAUMET-1998}.}
   \label{ldos_full}
\end{figure*}

We show in Fig.~\ref{ldos} the variations of normalized LDOS (ratio between the full LDOS and the LDOS in vacuum)
versus both the wavelength and the distance $d$ between the mirror and the center of the nanoparticle.
Figure~\ref{ldos_partielle} (red solid line) displays a section corresponding to $\lambda=400\,\textrm{nm}$ in Fig.~\ref{ldos}. 
We observe the well-known oscillations due to interferences between incident and reflected waves on the mirror~\cite{CHANCE-1978}. 
In the near-field regime corresponding to $d \ll  \lambda$, the relative
variations of the LDOS are on the order of $10\,\%$. This plots of the LDOS will be helpful in the qualitative analysis of
the variations of the generalized absorption cross-section. 

To study the influence of the mirror on $\sigma_a(\omega)$, we first consider an $s$-polarized illumination
(the incident plane wave has an electric field linearly polarized along the direction $y$). In this case, the induced 
electric dipole in the nanosphere is oriented along $y$. As a consequence, $\sigma_a(\omega)$
is independent on the direction of incidence (angle $\theta$ in Fig.~\ref{mirror}). 
In Fig.~\ref{sigma_s}, we represent the variations of  the normalized absorption cross-section $\sigma^n_a(\omega)$ 
(generalized absorption cross-section divided by free-space cross-section)
versus both the wavelength $\lambda$ and the distance $d$ between the mirror and the center of the nanoparticle.
The plasmon resonance, corresponding to $\lambda\simeq 360\,\textrm{nm}$, is visible for $d<100\,\textrm{nm}$.
Figure~\ref{ldos_sigma_s_p} displays a section view of figure~\ref{sigma_s} at $\lambda=361\,\textrm{nm}$. 
We observe oscillations of the absorption cross-section, as a clear signature of the influence of the mirror.
The relative variations are on the order of $5\,\%$ for distances $d$ between~$50$ and~$250\,\textrm{nm}$.
The behavior of $\sigma^n_a(\omega)$ can be compared to that of the partial LDOS $\rho_{yy}(\bm{r}_0,\omega)$
in Fig.~\ref{ldos_partielle}. The oscillations are in opposition, in agreement with the qualitative analysis presented
in section IV: An increase of the LDOS tends to decrease the absorption cross-section.

\begin{figure*}[!htbf]
   \centering
   \psfrag{d}[c]{$d\,(\mathrm{nm})$}
   \psfrag{l}[c]{$\lambda\,(\mathrm{nm})$}
   \psfrag{sh}[Bl]{$\sigma_a^n$}
   \subfigure[]{\label{sigma_s}\includegraphics[width=0.43\linewidth]{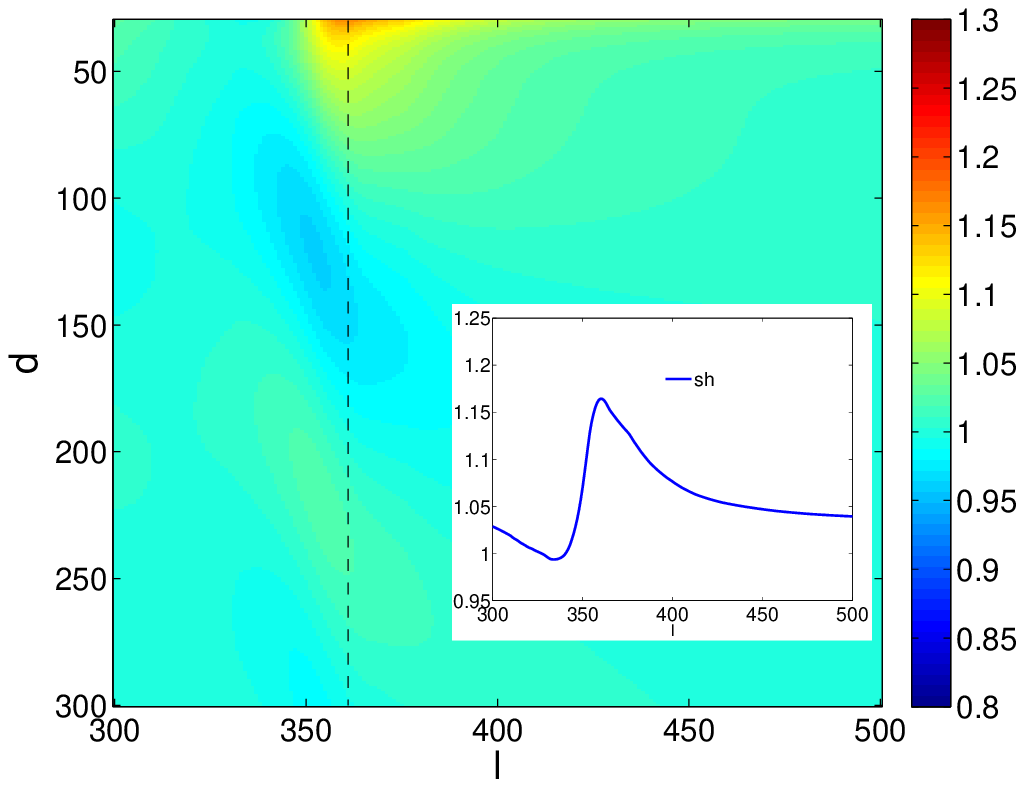}}
   \quad
   \psfrag{d}[c]{$d\,(\mathrm{nm})$}
   \psfrag{Lx}[c]{$\rho_{ii}(\bm{r}_0,\omega)$}
   \psfrag{s}[c]{$\sigma_a^n$}
   \psfrag{Lxh}[Bl]{$\rho_{yy}(\bm{r}_0,\omega)$}
   \psfrag{sh}[Bl]{$\sigma_a^n$}
   \subfigure[]{\label{ldos_sigma_s_p}\includegraphics[width=0.45\linewidth]{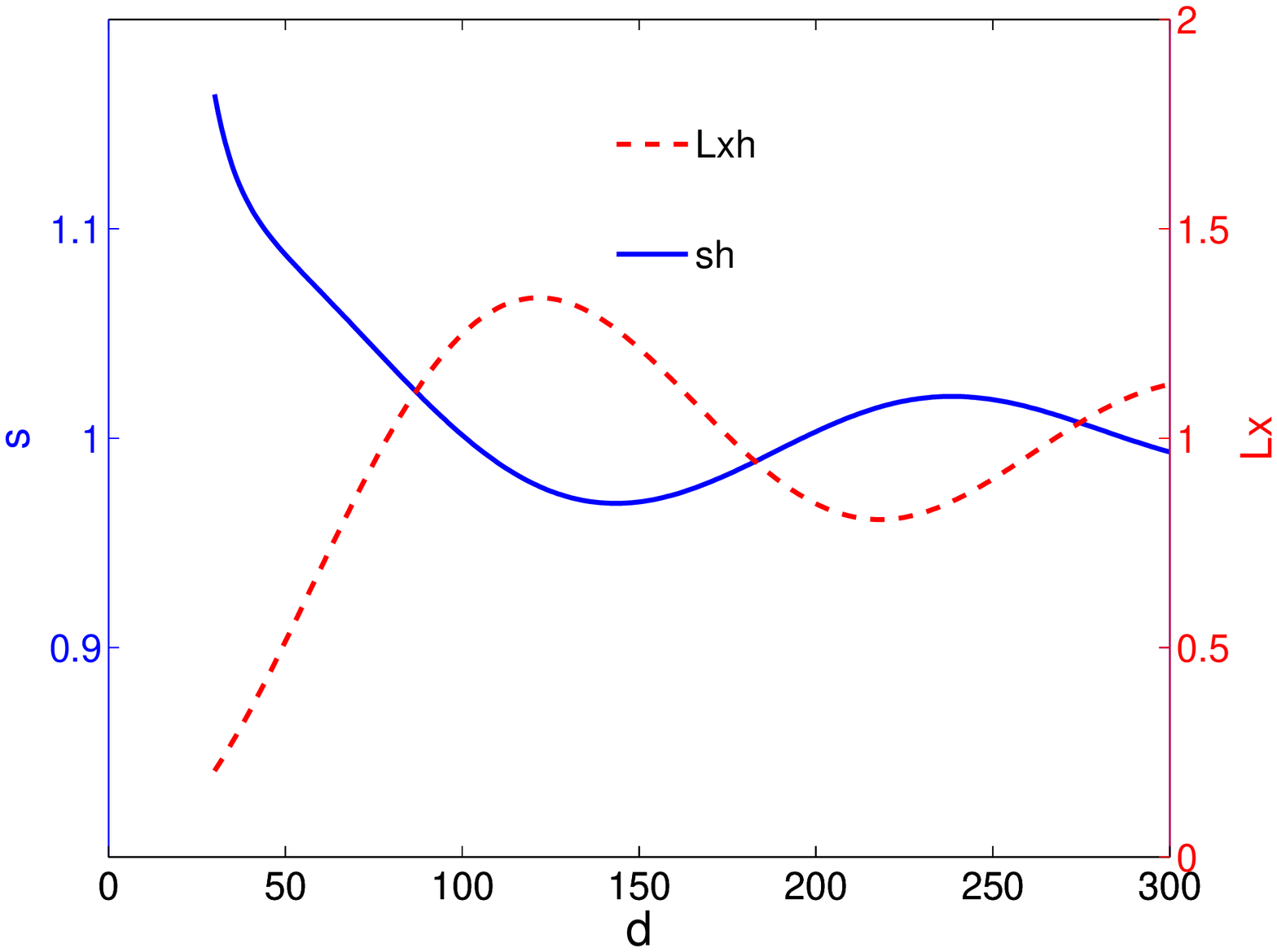}}
   \caption{\subref{sigma_s} Map of the normalized absorption cross-section $\sigma_a^n$ versus both the wavelength $\lambda$
   and the distance $d$ between the mirror and the nanosphere.
   The vertical dashed line is a guide to precise the wavelength at which panel~\subref{ldos_sigma_s_p} has been plotted.
   Inset: Plot of the absorption cross-section as a function of the wavelength for $d=30\,\textrm{nm}$.
   \subref{ldos_sigma_s_p} Plot of $\sigma_a^n$ and of the
   partial LDOS $\rho_{yy}(\bm{r}_0,\omega)$ versus the distance $d$ for $\lambda=361\,\textrm{nm}$. $s$-polarized illumination.}
   \label{sigma_s_full}
\end{figure*}

The case of an illumination with a $p$-polarized plane wave (i.e., with an electric field in the $x-z$ plane) can be analyzed
in a similar manner. In this case, the electric dipole induced in the nanosphere depends on the direction of incidence $\theta$. 
We have chosen $\theta=45^{\textrm{o}}$ for the sake of illustration (note that $\theta=0$ would lead to the same behavior as that observed
with $s$-polarized illumination).
Figures~\ref{sigma_p_oblique} and \ref{ldos_sigma_s_p_oblique} are the same as Figs.~\ref{sigma_s} and \ref{ldos_sigma_s_p}, but
for $p$-polarized illumination.
We observe a similar  behavior of $\sigma_a^n$, with oscillations corresponding to relative variations of a few percent. Close to
resonance, and for $d=30\,\textrm{nm}$ (which corresponds to strong nanoparticle-mirror interaction in this simple system),
 the relative variation of the absorption cross-section is of the order of $20\,\%$.
Finally, let us note that significant changes are observed in the near-field regime only. For $d> \lambda/2$, the influence of interactions
with the environment remains weak. 

\begin{figure*}[!htbf]
   \centering
   \psfrag{d}[c]{$d\,(\mathrm{nm})$}
   \psfrag{l}[c]{$\lambda\,(\mathrm{nm})$}
   \psfrag{sh}[Bl]{$\sigma_a^n$}
   \subfigure[]{\label{sigma_p_oblique}\includegraphics[width=0.43\linewidth]{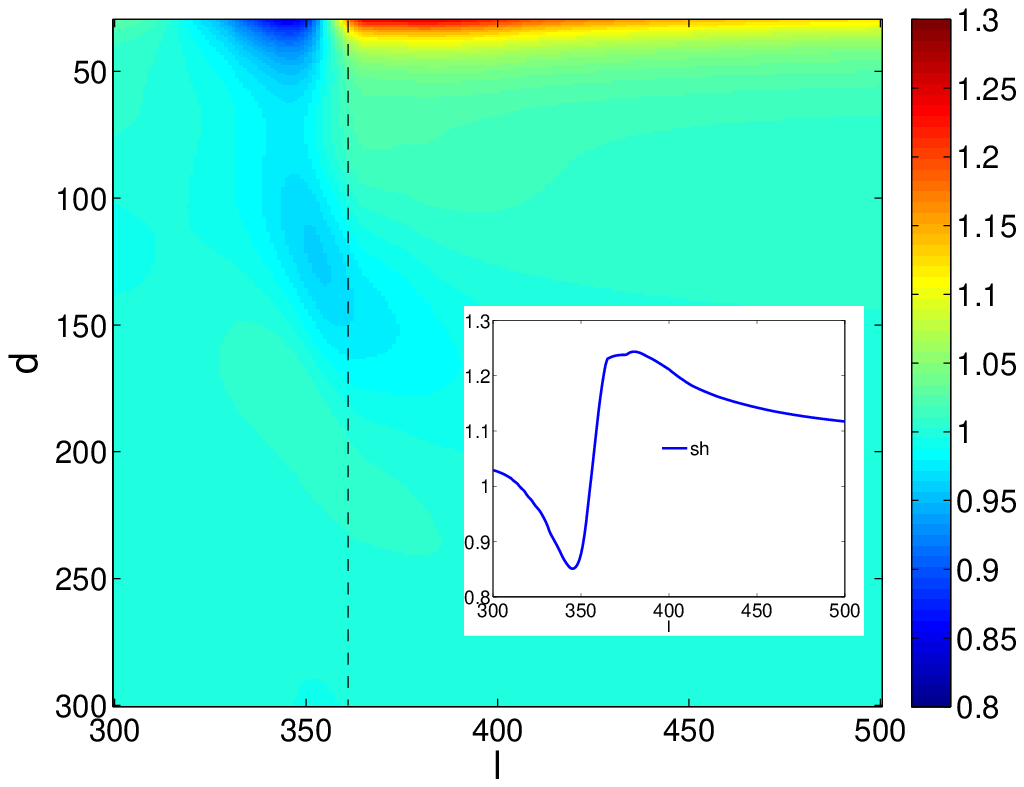}}
   \quad
   \psfrag{d}[c]{$d\,(\mathrm{nm})$}
   \psfrag{Lx}[c]{$\rho_{ii}(\bm{r}_0,\omega)$}
   \psfrag{s}[c]{$\sigma_a^n$}
   \psfrag{Lxh}[Bl]{$\rho_{xx}(\bm{r}_0,\omega)$}
   \psfrag{Lzh}[Bl]{$\rho_{zz}(\bm{r}_0,\omega)$}
   \psfrag{sh}[Bl]{$\sigma_a^n$}
   \subfigure[]{\label{ldos_sigma_s_p_oblique}\includegraphics[width=0.45\linewidth]{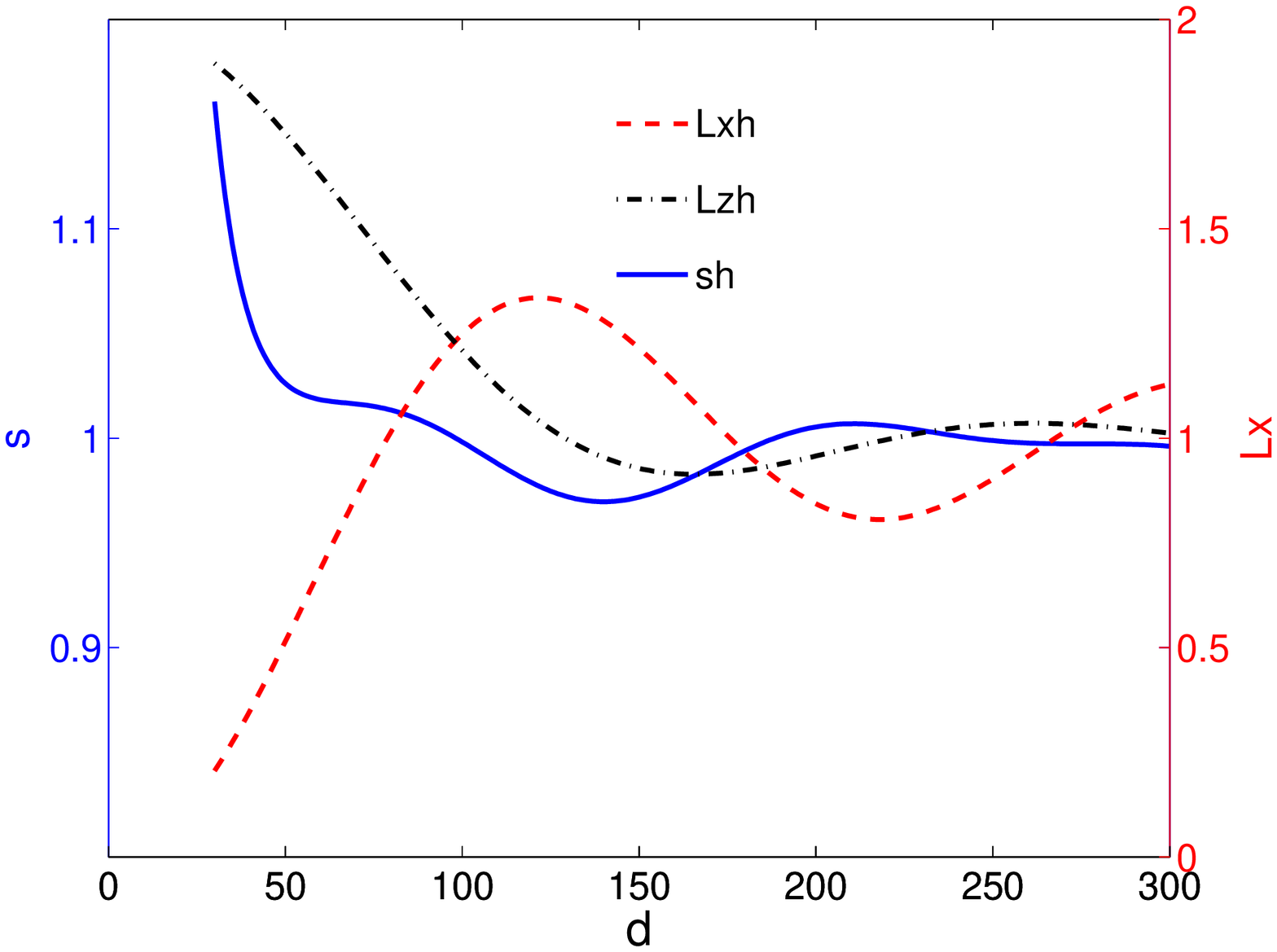}}
   \caption{\subref{sigma_p_oblique} Same as Fig.~\ref{sigma_s_full} for an illumination with a $p$-polarized plane wave,
   at an angle of incidence $\theta=45^{\textrm{o}}$. Both the partial LDOS $\rho_{xx}(\bm{r}_0,\omega)$ and $\rho_{zz}(\bm{r}_0,\omega)$
   are displayed in~\subref{ldos_sigma_s_p_oblique} since both influence the generalized cross-section in this situation.}
   \label{sigma_p_oblique_full}
\end{figure*}

\section{Conclusion}

We have described the influence of the environment on the absorption cross-section of an optical dipole antenna or a nanoparticle,
based on a rigorous framework.  We have derived a generalized form of the absorption cross-section, based on the only assumption
that  the electric field is uniform inside the antenna (electric dipole approximation). 
In the case of a resonant nanoparticle (plasmon resonance), we have analyzed qualitatively the role of the environment on 
the resonance frequency and on the linewidth. In particular, we have identified the role of the photonic LDOS, and shown
that an increase of the LDOS results in a reduction of the generalized absorption cross-section. These effects have been illustrated
on the simple example of a metallic nanoparticle interacting with a perfect mirror. 
In the field of optical nanoantennas, these results could be exploited along two directions. First, engineering the LDOS around an optical 
nanoantenna could allow some control of both the resonance frequency, and more interestingly on the level of absorption. Since
high absorption remains a serious drawbacks of metallic nanoantenna, it might be possible to reduce absorption by an appropriate
structuration of the environment. Second, measuring changes in the resonance lineshape  of a metallic nanoparticle,
as performed, e.g., in Ref.~\cite{SANDOGHDAR-2005-1}, should allow a direct mapping of the LDOS, without using fluorescent emitters.
Finally, let us comment on the possibility of measuring $\sigma_a$ \emph{in situ}. A potential method could be based on
photothermal detection, in which a probe beam probes the temperature increase of the nanoparticle due to absorption. Such
methods already offer the possibility of sensitive detection of nanoparticles in complex 
environments~\cite{BOYER-2002,LOUNIS-2002,LOUNIS-2006,FOURNIER-2010}.

This work was supported by the EU Project \emph{Nanomagma} under Contract No.~NMP3-SL-2008-214107.
E.C. acknowledges a doctoral grant from the French DGA.

\appendix

\section{Proof of the relation $\im[\tens{\alpha}_0][\tens{\alpha}_0\tens{\alpha}_0^*]^{-1}=\im\epsilon(\omega)/\left[V|\epsilon(\omega)-1|^2\right]\tens{I}$}
\label{proof}

In this appendix, we give a proof of the relation $\im[\tens{\alpha}_0][\tens{\alpha}_0\tens{\alpha}_0^*]^{-1}=\im\epsilon(\omega)/\left[V|\epsilon(\omega)-1|^2\right]\tens{I}$. The quasi-static polarizability is given by Eq.~(\ref{pola0}):
\begin{align}\nonumber
   \tens{\alpha}_0(\omega)=V[\epsilon(\omega)-1]\left\{\tens{I}+[\epsilon(\omega)-1]\tens{L}\right\}^{-1}.
\end{align}
Multiplying the previous expression on the left by $\left\{\tens{I}+[\epsilon^*(\omega)-1]\tens{L}\right\}\left\{\tens{I}+[\epsilon^*(\omega)-1]\tens{L}\right\}^{-1}$, we end-up with
\begin{align}\nonumber
   \im\tens{\alpha}_0(\omega)= & V\im\epsilon(\omega)
\\\label{alpha1} & \times
   \left\{\tens{I}+2\re[\epsilon(\omega)-1]\tens{L}+|\epsilon(\omega)-1|^2\tens{L}^2\right\}^{-1}.
\end{align}
However, we also have
\begin{align}\nonumber
   \tens{\alpha}_0\tens{\alpha}_0^*= & V^2|\epsilon(\omega)-1|^2
\\\label{alpha2} & \times
   \left\{\tens{I}+2\re[\epsilon(\omega)-1]\tens{L}+|\epsilon(\omega)-1|^2\tens{L}^2\right\}^{-1}.
\end{align}
Using Eqs.~(\ref{alpha1}) and~(\ref{alpha2}), we obtain the following relationship:
\begin{align}\nonumber
   \im[\tens{\alpha}_0][\tens{\alpha}_0\tens{\alpha}_0^*]^{-1}=\frac{\im\epsilon(\omega)}{V|\epsilon(\omega)-1|^2}\tens{I}
\end{align}
which concludes the proof.

\section{Derivation of the scalar form of the absorption cross-section at resonance}
\label{resonant}

To derive Eq.~(\ref{absorption_resonant}) we first compute the derivative of Eq.~(\ref{absorption_scalaire}) with
respect to $\omega$:
\begin{align}\nonumber
   \frac{\ud\sigma_a}{\ud\omega} & (\omega) = \frac{6V\gamma \omega\omega_0^2}{c}
      \left[
         \{\Omega_{\textrm{eff}}^2(\omega)-\omega^2\}^2
      \right.
\\\nonumber &
         -2\omega\{\Omega_{\textrm{eff}}'(\omega)\Omega_{\textrm{eff}}(\omega)-\omega\}\{\Omega_{\textrm{eff}}^2(\omega)-\omega^2\}
\\\nonumber &
      \left.
         -\omega^3\gamma_{\textrm{eff}}(\omega)'\gamma_{\textrm{eff}}(\omega)
      \right]
      /
      \left\{[\Omega_{\textrm{eff}}^2(\omega)-\omega^2]^2+\omega^2\gamma_{\textrm{eff}}^2(\omega)\right\}^2.
\end{align}
The resonance frequency $\omega_a$ is defined by $\ud\sigma_a(\omega_a)/\ud\omega=0$. This gives us a relation
satisfied by $\omega_a$. By factorizing by $\Omega_{\textrm{eff}}^2(\omega_a)-\omega_a^2$, and by putting this
relation into Eq.~(\ref{absorption_scalaire}) we end up with
\begin{align}\nonumber
   \sigma_a & (\omega_a) = \frac{3\gamma V\omega_0^2}{c} \left[\gamma_{\textrm{eff}}^2(\omega_a)\left\{
      1
   \vphantom{\frac{\omega_a^4}{\omega_a^4}}
   \right.\right.
\\\nonumber &
   \left.\left.
      +\frac{\omega_a^4}{[\Omega_{\textrm{eff}}^2(\omega_a)-2\omega_a\Omega_{\textrm{eff}}'(\omega_a)\Omega_{\textrm{eff}}(\omega_a)
      +\omega_a^2]^2}\gamma_{\textrm{eff}}'^2(\omega_a)
   \right\}\right]^{-1}
\end{align}
which concludes the derivation of Eq.~(\ref{absorption_resonant}).

\section{Computation of the exciting field and of the Green function of the perfect mirror system}
\label{computation}

The computation of the modified absorption cross-section for a metallic nanosphere close to a perfect mirror
requires the computation of the exciting field and of the Green tensor of the system composed by the mirror only.
In the case of a $s$-polarised incident field given by
\begin{align}\nonumber
   \e_{\textrm{inc}}^s(\bm{r},\omega)=E_0\exp[ik_x x+ik_z z]\bm{e}_y
\end{align}
where the incident wave-vector is $\bm{k}=k_0(-\sin\theta,0,-\cos\theta)$, the exciting field at the position $\bm{r}_0$ of the nanosphere
is simply given by the superposition of the incident and reflected fields:
\begin{align}\nonumber
   \e_{\textrm{ext}}^s(\bm{r}_0,\omega)=-2iE_0\sin[k_0\cos\theta d]\bm{e}_y.
\end{align}
In the same way, in the case of a $p$-polarised incident field given by
\begin{align}\nonumber
   \e_{\textrm{inc}}^p(\bm{r},\omega)=E_0\exp[ik_x x+ik_z z]\{\cos\theta\bm{e}_x+\sin\theta\bm{e}_z\},
\end{align}
the exciting field reads
\begin{align}\nonumber
   \e_{\textrm{ext}}^p(\bm{r}_0,\omega)= & -2iE_0\cos\theta\sin[k_0\cos\theta d]\bm{e}_x
\\\nonumber &
   +2E_0\sin\theta\cos[k_0\cos\theta d]\bm{e}_z.
\end{align}

In order to express the Green function, we use the dipole image method. If the system consisting of the perfect mirror at $z=0$
is illuminated by a source dipole $\bm{p}=(p_x,p_y,p_z)$ at position $\bm{r}_0=(0,0,z_0)$, the effect of the mirror can be replaced by 
the radiation of an image dipole $\bm{p}'=(-p_x,-p_y,p_z)$ placed at position $\bm{r}_0'=(0,0,-z_0)$. 
This allows us to compute the Green function of the system in a simple manner:
\begin{align}\nonumber
   \tens{G}(\bm{r},\bm{r}_0,\omega)\bm{p}=\tens{G}_0(\bm{r},\bm{r}_0,\omega)\bm{p}+\tens{G}_0(\bm{r},\bm{r}_0',\omega)\bm{p}'.
\end{align}
In this expression, $\tens{G}_0$ is the Green tensor in vacuum given by
\begin{align}\nonumber
   \tens{G}_0(\bm{r},\bm{r}_0,\omega)= & \operatorname{PV}\left\{
         \left[\tens{I}-\bm{u}\otimes\bm{u}+\frac{ik_0R-1}{k_0^2R^2}\left(\tens{I}-3\bm{u}\otimes\bm{u}\right)\right]
      \right.
\\\nonumber & \hphantom{\operatorname{PV}\{}
      \left.
         \times\frac{\exp\left[ik_0R\right]}{4\pi R}
      \right\}
      -\frac{\delta\left(\bm{R}\right)}{3k_0^2}\tens{I}
\end{align}
with $\bm{R}=\bm{r}-\bm{r}_0$ and $\bm{u}=\bm{R}/R$. $\operatorname{PV}$ denotes the principal value operator. At the position
of the nanosphere (i.e. $\bm{r}=\bm{r}_0$), the singularity
part of the Green tensor [needed to compute the quasi-static polarisability $\alpha_0$ given by Eq.~(\ref{pola0})] 
is simply  $\tens{L}=\tens{I}/3$.
The regular part [needed to compute the polarisability $\alpha$ given by Eq.~(\ref{pola_final})] is given by
\begin{align}\nonumber
   \tens{G}_{\textrm{reg}}(\bm{r}_0,\bm{r}_0,\omega)\bm{p}=
   \tens{G}_{0,\textrm{reg}}(\bm{r}_0,\bm{r}_0,\omega)\bm{p}+
   \tens{G}_0(\bm{r}_0,\bm{r}_0',\omega)\bm{p}'.
\end{align}


\end{document}